\documentclass{article}
\usepackage{spconf,amsmath,graphicx}
\usepackage{subfigure,color}
\usepackage{algorithm,algorithmic}
\usepackage[percent]{overpic}

\title{Quarter Laplacian Filter for Edge Aware Image Processing}
\name{Yuanhao Gong\thanks{Thanks to National Natural Science Foundation for funding (61907031).}, Wenming Tang, Lebin Zhou, Lantao Yu, Guoping Qiu}
\address{College of Electronics and Information Engineering, Shenzhen University, China, gong@szu.edu.cn}
\begin{document}
%
\maketitle
\begin{abstract}
This paper presents a quarter Laplacian filter that can preserve corners and edges during image smoothing. Its support region is $2\times2$, which is smaller than the $3\times3$ support region of Laplacian filter. Thus, it is more local. Moreover, this filter can be implemented via the classical box filter, leading to high performance for real time applications. Finally, we show its edge preserving property in several image processing tasks, including image smoothing, texture enhancement, and low-light image enhancement. The proposed filter can be adopted in a wide range of image processing applications. 
\end{abstract}
\section{Introduction}
\label{sec:intro}
Image processing covers a wide range of tasks, including denoising, segmentation, reconstruction, inpainting, etc. Among these applications, image smoothing is a fundamental task. It can be used to remove details of the image. It can also be used to remove the noise in the image. Its applications include depth estimation, optical flow estimation, stereo vision, surface reconstruction, object detection, etc.

There are various image smoothing methods that have been developed in the last decades. The classical box filter simply takes the average value in one pixel's neighborhood. The well-known Gaussian filter exploits the Gaussian-determined weights as the kernel to obtain a weighted average. Another well-known filter is the Laplacian filter, which originates from the standard diffusion equation. 

\subsection{Edge Preserving Image Filters}
In image smoothing, it is preferred to preserve the large gradient while remove the small gradient. Such property is called edge preserving. The removed details are usually called texture while the smoothed result is called structure.

There are many filters that have the edge preserving property. The well-known bilateral filter introduces weights that not only depend on spatial coordinates but also on the image intensities \cite{Tomasi:1998}. Another popular edge preserving filter is the guided image filter \cite{he2010guided}, which assumes that each patch in the input image is a linear function of the corresponding patch in the guided image. 
\subsection{Deep Learning Based Image Filtering}
The edge preserving property can also be achieved by neural networks, for example, convolution neural networks \cite{Xu2015,2017Deep,Chen2017}. Various neural network architectures have been developed for different image processing tasks. 

However, these convolution neural networks can not be adopted for high resolution images, because they usually require a large mount of computer memory to perform the computation. Due to this limitation, they can not be used for real time applications.

Another problem for neural network methods is that they might lead to bizarre artifacts. This is because these networks are not currently interpretable and they are without any theoretical guarantees. Such issue hampers their applications in practical image processing tasks.

Thus, simple filter based image smoothing algorithms are still needed in practical applications, especially for real time scenarios and other cases that require high reliability. 
\subsection{Laplacian Filter}
One of the classical simple filters is the Laplacian filter. It is a high pass filter from signal processing point of view. Its name comes from the second order diffusion operator, Laplacian operator, in mathematics. It is an isotropic diffusion operator and thus can not preserve edges. 
\subsection{One-sided Window Filter}
In our previous work \cite{gong:phd,Yin2019,gong:cf,Gong2018}, we put several one-sided windows around the pixel to be processed. Such side windows allow discontinuity in the results. Therefore, the resulting images can have sharp edges. This idea is general and can be combined with Gaussian filter, box filter, bilateral filter, guided filter, etc. The one-sided window version preserves edges better than the original method. 
\subsection{Our Contributions}
In this paper, we further reduce the one-sided windows to quarter windows. More specifically, we propose a quarter window Laplacian filter, which has following properties:
\begin{itemize}
	\item it can preserve corners and edges;
	\item it has $2\times2$ support region, which is smaller than $3\times3$ in Laplacian filter;
	\item it has high performance and can be implemented by a simple box filter.
\end{itemize}

\section{Quarter Laplacian Filter}
The classical Laplacian operator is an isotropic diffusion operator. In this paper, we propose to perform this operator on four quarter windows, leading to an anisotropic diffusion (in other words, edge preserving). The proposed quarter Laplacian filter can preserve corners and edges. It can be implemented by the box filter and thus achieve high performance.
\subsection{Discrete Laplacian Operator}
In the continuous case, the Laplacian operator comes from the standard diffusion process
\begin{equation}
	\frac{\mathrm{d}U(\vec{x})}{\mathrm{d}t}=c\Delta U(\vec{x})\,
\end{equation} where $c$ is a diffusion coefficient scalar, $t$ is the (pseudo) time, $\Delta$ is the standard Laplacian operator, $U(\vec{x})$ is the unknown function, and $\vec{x}$ is the spatial coordinate.

For discrete digital images, this equation must be discretized. More specifically, the discrete diffusion becomes
\begin{equation}
	\label{eq:diffusion}
	U^{t+1}(x_i,y_j)=U^{t}(x_i,y_j)+c\Delta U^t(x_i,y_j)\,,
\end{equation} where $i,j$ are integer indexes. For convenience, we will drop $(x_i,y_j)$. The classical Laplacian operator is usually carried out by one of the following convolution kernels
\begin{equation}
	\label{eq:kernels}
	\small
	\begin{pmatrix} 0 & \frac{1}{4} & 0 \\ \frac{1}{4} & -1&\frac{1}{4} \\
		0 & \frac{1}{4} & 0
	\end{pmatrix},	
	\begin{pmatrix} \frac{-1}{16} & \frac{5}{16} & \frac{-1}{16} \\ \frac{5}{16} & -1&\frac{5}{16} \\
		\frac{-1}{16} & \frac{5}{16} & \frac{-1}{16}
	\end{pmatrix},
	\begin{pmatrix} \frac{1}{12} & \frac{1}{6} & \frac{1}{12} \\ \frac{1}{6} & -1&\frac{1}{6} \\
		\frac{1}{12} & \frac{1}{6} & \frac{1}{12}
	\end{pmatrix}\,.
\end{equation}

\begin{figure}
	\centering
	\subfigure[left in \eqref{eq:kernels}]{\includegraphics[width=0.32\linewidth]{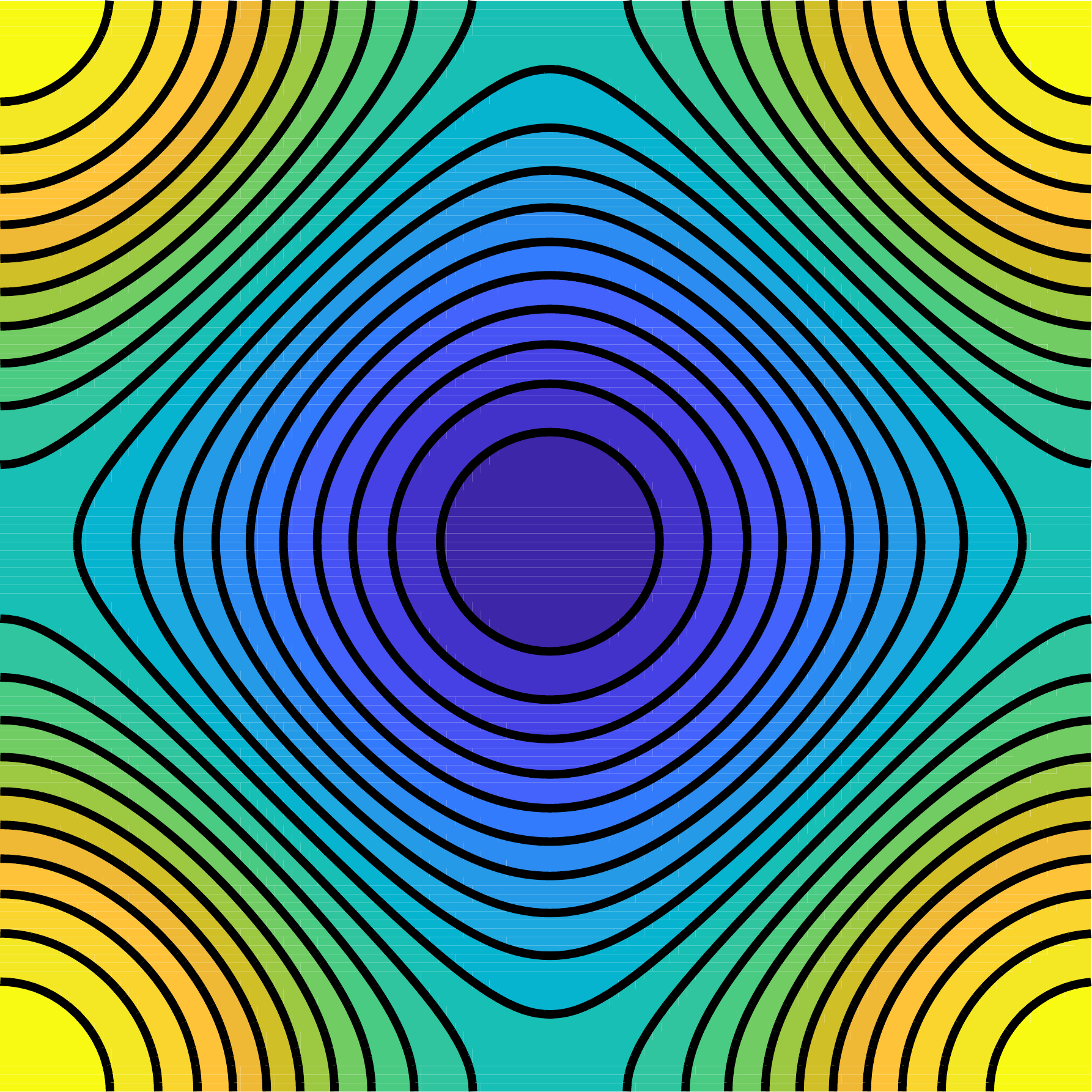}}
	\subfigure[middle in \eqref{eq:kernels}]{\includegraphics[width=0.32\linewidth]{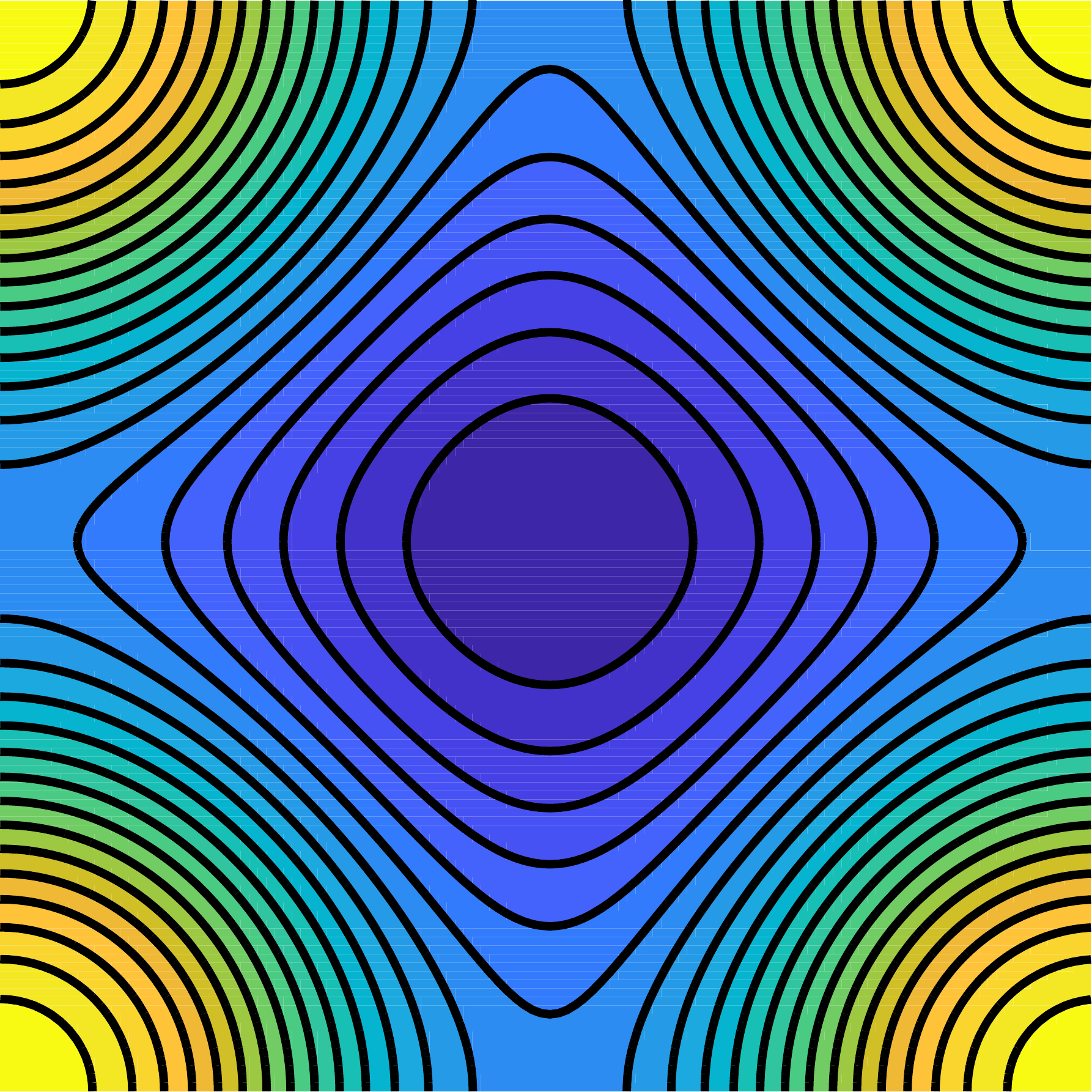}}
	\subfigure[right in \eqref{eq:kernels}]{\includegraphics[width=0.32\linewidth]{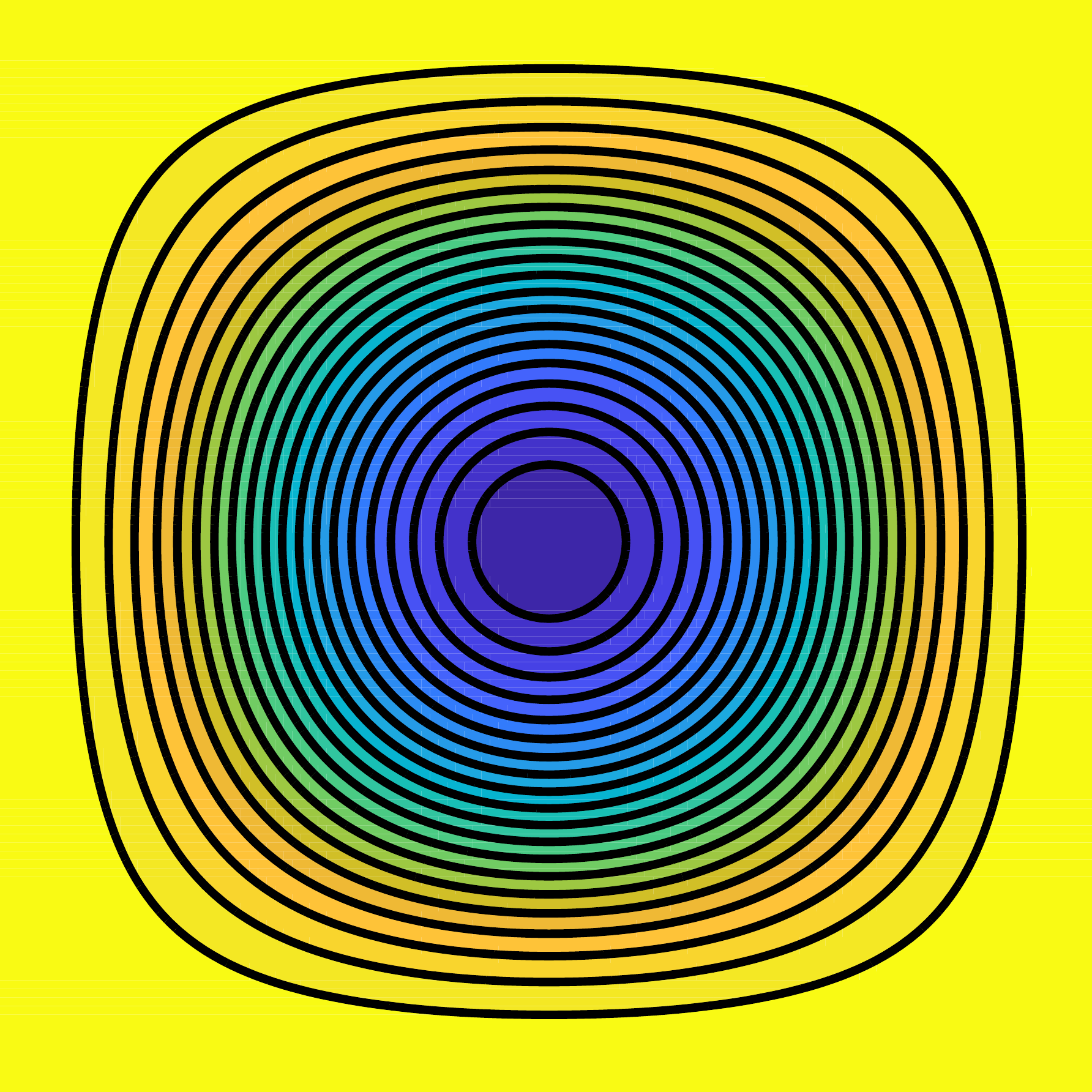}}
	\caption{Spectral analysis of discrete Laplacian operators}
	\label{fig:ks}
\end{figure}

The left kernel in \eqref{eq:kernels} is well-known in image processing. The middle kernel in \eqref{eq:kernels} is developed in \cite{gong:phd,gong:cf}. The right kernel in \eqref{eq:kernels} is from \cite{isoLaplace:1999}. Their Fourier spectrum is shown in Fig.~\ref{fig:ks}. The right kernel is the most isotropic one. Thus, we take it as the discrete Laplacian operator in this paper.

\subsection{Quarter Windows}
\begin{figure}
	\centering
	\subfigure[four quarter windows for the center location (red dot)]{\includegraphics[width=0.99\linewidth]{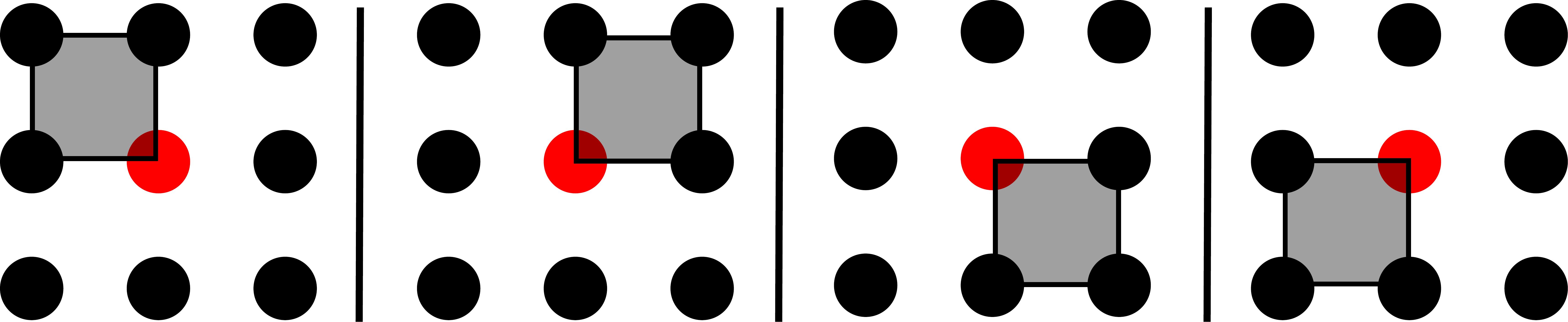}}
	\subfigure[four half windows from our previous work~\cite{gong:Bernstein}]{\includegraphics[width=0.99\linewidth]{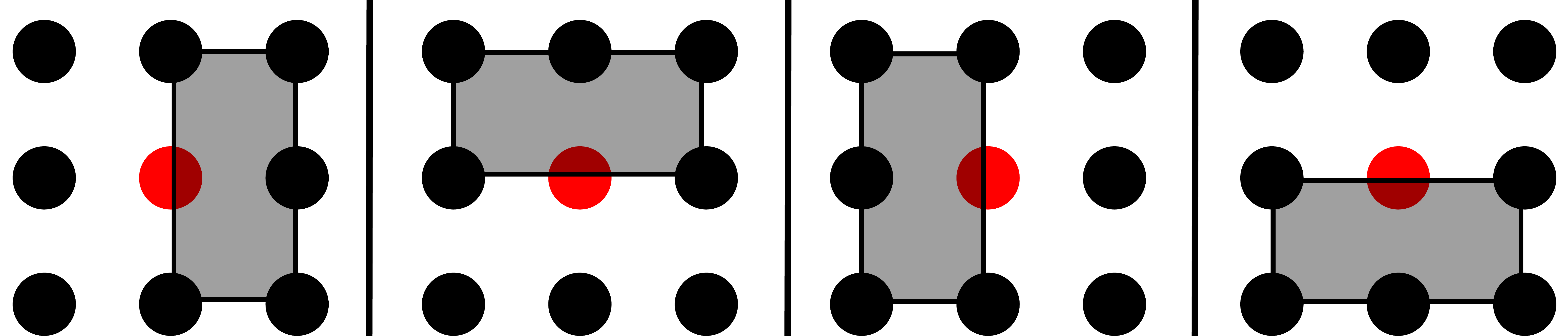}}
	\caption{Quarter windows and half windows for the center location (red dot). In this paper, we only use the quarter windows.}
	\label{fig:windows}
\end{figure}
Inspired by our previous works~\cite{gong:Bernstein,GONG2019329,gong2013a}, we use quarter windows as shown in top row of Fig.~\ref{fig:windows}. There are several reasons that quarter windows are preferred. First, the quarter windows have smaller support region ($2\times2$ in this case) than the half windows ($2\times3$ or $3\times2$). Smaller support regions indicate more local geometric information. Second, the quarter window Laplacian filter can be implemented by box filter, leading to high performance (Section~\ref{sec:fast}). 

\subsection{Quarter Laplacian Filter}
\begin{figure}
	\centering
	\subfigure[input]{\includegraphics[width=0.32\linewidth]{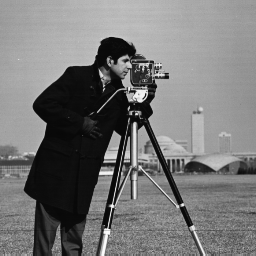}}
	\subfigure[Laplacian]{\includegraphics[width=0.32\linewidth]{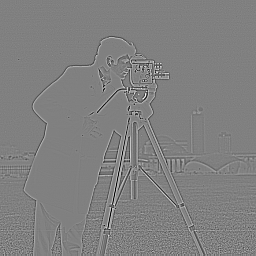}}
	\subfigure[QuarterLap $d_m$]{\includegraphics[width=0.32\linewidth]{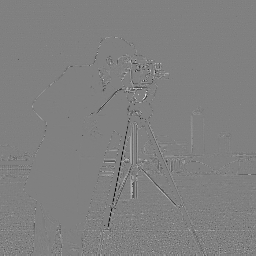}}
	
	\subfigure[$d_1$]{\includegraphics[width=0.24\linewidth]{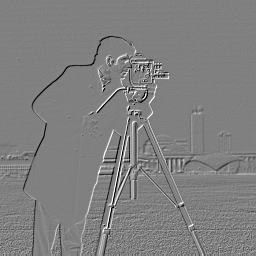}}
	\subfigure[$d_2$]{\includegraphics[width=0.24\linewidth]{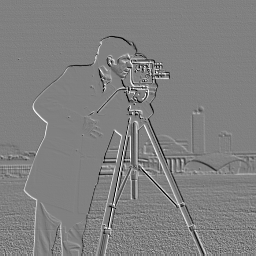}}
	\subfigure[$d_3$]{\includegraphics[width=0.24\linewidth]{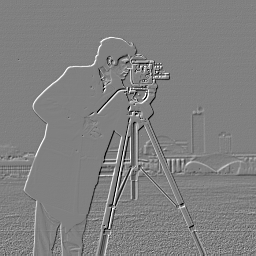}}
	\subfigure[$d_4$]{\includegraphics[width=0.24\linewidth]{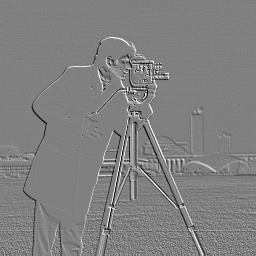}}
	\caption{Input (a), Laplacian feature map (b), Quarter Laplacian feature map (c) and its intermediate features (d, e, f, g). All feature maps are added 128 for better visualization.}
	\label{fig:maps}
\end{figure}

Combining these quarter windows with the isotropic Laplacian operator in \eqref{eq:kernels}, we obtain following convolution kernels
\begin{alignat*}{2}
		\label{eq:qk}
	k_1 &= \left[ \begin{array}{ccc} \frac{1}{3} & \frac{1}{3} & 0 \\
		\frac{1}{3} & -1 & 0 \\
		0 & 0 & 0
	\end{array} \right]\!,\,&
	k_2 &= \left[ \begin{array}{ccc} 0 & \frac{1}{3} & \frac{1}{3} \\
		0 & -1 & \frac{1}{3} \\
		0 & 0 &0 
	\end{array} \right] \,,\\
	k_3 &= \left[ \begin{array}{ccc} 0 & 0 & 0 \\
		0 & -1 & \frac{1}{3} \\
		0 & \frac{1}{3} & \frac{1}{3}
	\end{array} \right]\!,\,&
	k_4 &= \left[ \begin{array}{ccc} 0 & 0 & 0 \\
		\frac{1}{3} & -1 & 0 \\
		\frac{1}{3} & \frac{1}{3} &0
	\end{array} \right] \,.
\end{alignat*}
From these kernels, we can compute four feature maps
\begin{equation}
	\label{eq:conv}
	d_i=k_i\ast U\,,~\forall i=1,..,4\,,
\end{equation} where $\ast$ is the convolution operator.
Then, only one feature map $d_{m(x,y)}(x,y)$ is selected, where
\begin{equation}
	\label{eq:act}
	m(x,y)=\arg\min_i\{|d_i(x,y)|;\,i=1,..,4\}\,.
\end{equation} Such selection provides the non-linearity for our filter. This $d_{m(x,y)}(x,y)$ is the result of the quarter Laplacian filtering. This filter is summarized in Algorithm~\ref{algo:HW}. One example of this quarter Laplacian filter is shown in Fig.~\ref{fig:maps}.

\begin{algorithm}[!htb]
	\caption{Quarter Laplacian Filter}
	\label{algo:HW}
	\begin{algorithmic}
		\REQUIRE input image $U(x,y)$
		\STATE  $d_i(x,y)=k_i\ast U(x,y),\,i=1,..,4$
		\STATE $d_{m(x,y)}(x,y)$ where $m=\arg\min_i\{|d_i(x,y)|;\,i=1,..,4\}$
		\ENSURE $d_m(x,y)$
	\end{algorithmic}
\end{algorithm}

\subsection{Fast Implementation}
\label{sec:fast}
One advantage of adopting quarter windows for Laplacian operator is that the weights in kernels $k_i$ become constant ($\frac{1}{3}$ in this case). Therefore, the quarter Laplacian filter can be implemented by box filter, leading to high performance.

Take the kernel $k_1$ as an example (other kernels can be analyzed similarly). This kernel can be rewritten as
\begin{equation}
	k_1=\begin{bmatrix} \frac{1}{3} & \frac{1}{3} & 0 \\ \frac{1}{3} & -1&0 \\
		0 & 0 & 0
	\end{bmatrix}=\frac{1}{3}
\begin{bmatrix} 1 & 1 & 0 \\ 1 & 1&0 \\
	0 & 0 & 0
\end{bmatrix}-
\begin{bmatrix} 0 & 0 & 0 \\
	0 & \frac{4}{3} & 0\\
	0 & 0 & 0
\end{bmatrix}.
\end{equation}
Therefore, all the convolutions with $k_i$ can be carried out by a box or average filter. 

\begin{figure}[t]
	\centering
	\includegraphics[width=0.35\linewidth]{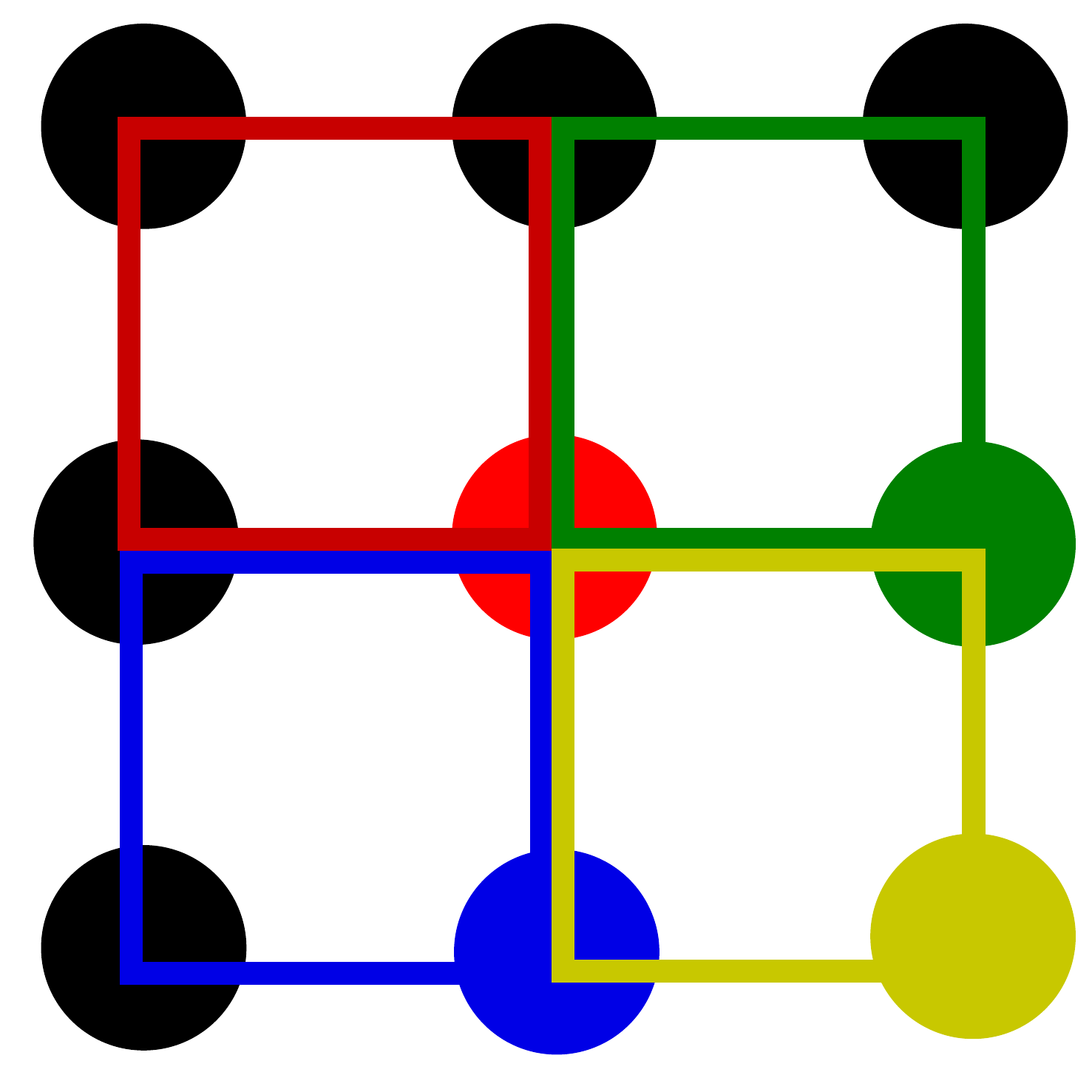}
	\caption{Overlapped support regions for different locations can be used to reduce the computation. For example, the bottom-right region for the red dot is exactly the upper left region for the yellow dot.}
	\label{fig:fast}
\end{figure}

Moreover, thanks to the overlapped support regions, only one convolution is needed, instead of four. Specifically, as shown in Fig.~\ref{fig:fast}, the upper left support region of the yellow dot location is exactly the right bottom support region of the red dot location. As a result, only the upper left region for each pixel is needed. With this technique, the proposed filter has almost the same runtime as the original Laplacian filter.

\begin{figure}[!bh]
	\centering
	\subfigure[Laplacian $t=10$]{\includegraphics[width=0.32\linewidth]{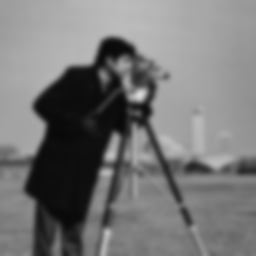}}
	\subfigure[Laplacian $t=100$]{\includegraphics[width=0.32\linewidth]{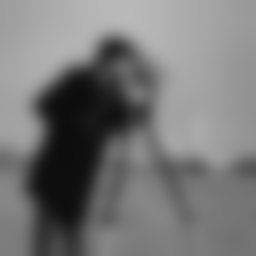}}
	\subfigure[Laplacian $t=1000$]{\includegraphics[width=0.32\linewidth]{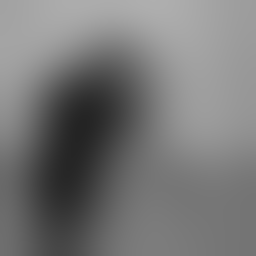}}
	
	\subfigure[Quarter $t=10$]{\includegraphics[width=0.32\linewidth]{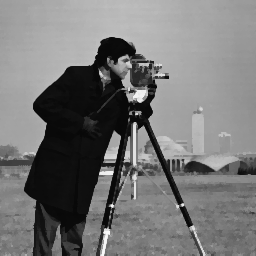}}
	\subfigure[Quarter $t=100$]{\includegraphics[width=0.32\linewidth]{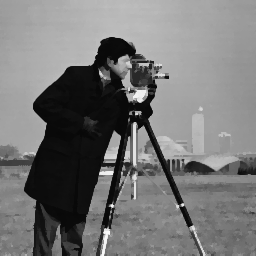}}
	\subfigure[Quarter $t=1000$]{\includegraphics[width=0.32\linewidth]{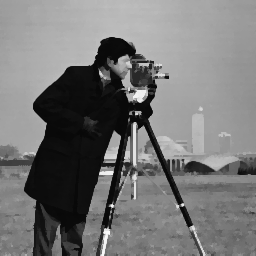}}
	\subfigure[the 128th row profiles for the first 20 iterations]{\includegraphics[width=\linewidth]{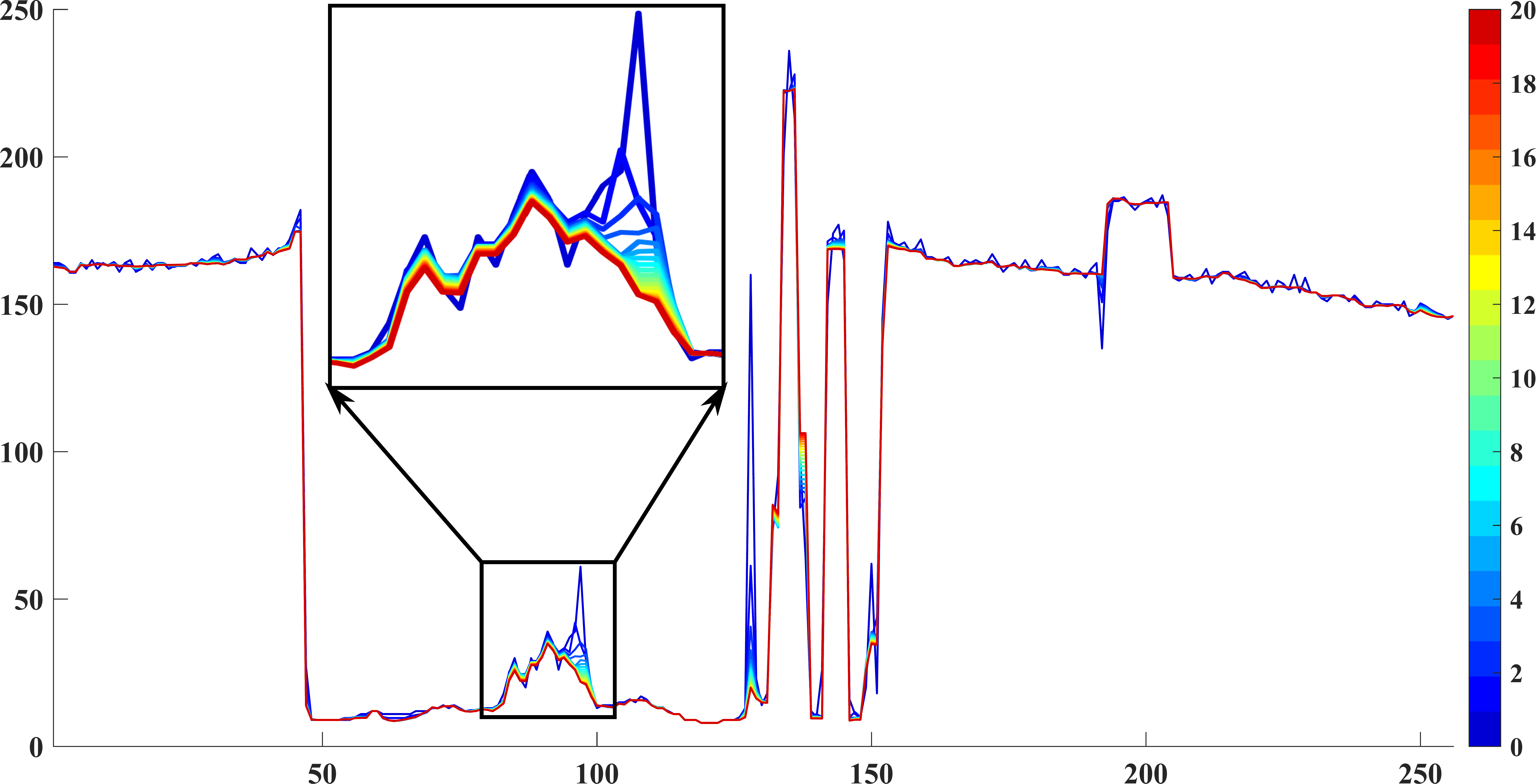}}
	\caption{First row: diffusion by the Laplacian filter; second row: diffusion by the quarter Laplacian filter; third row: line profiles for the first 20 iterations in quarter Laplacian filter.}
	\label{fig:diff}
\end{figure}
\subsection{Edge Preserving}
In the diffusion process \eqref{eq:diffusion}, we replace the Laplacian filter with the quarter Laplacian filter
\begin{equation}
	U^{t+1}=U^t+\mathrm{QuarterLaplacianFilter}(U^t)\,.
\end{equation} Then an edge preserving anisotropic diffusion process is obtained. Such difference is shown in Fig.~\ref{fig:diff}. Quarter Laplacian filter preserves corners and edges while Laplacian filter does not. The 128th row profiles for first 20 iterations are shown in Fig.~\ref{fig:diff} (g). A fragment is zoomed for better visualization. After 10 iterations, the result does not obviously change. In the rest of this paper, we set 10 as the default iteration number.

\begin{figure*}
	\centering
	{\includegraphics[width=0.16\linewidth]{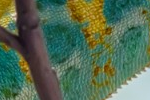}}\hspace*{-1pt}
	{\includegraphics[width=0.16\linewidth]{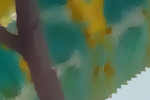}}\hspace*{-1pt}
	{\includegraphics[width=0.16\linewidth]{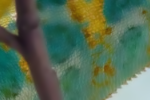}}\hspace*{-1pt}
	{\includegraphics[width=0.16\linewidth]{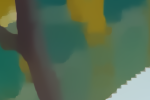}}\hspace*{-1pt}
	{\includegraphics[width=0.16\linewidth]{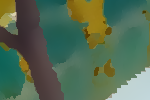}}\hspace*{-1pt}
	{\includegraphics[width=0.16\linewidth]{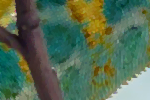}}\\
	\vspace{2pt}

	{\includegraphics[width=0.16\linewidth]{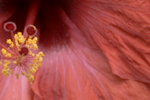}}\hspace*{-1pt}
	{\includegraphics[width=0.16\linewidth]{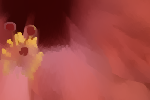}}\hspace*{-1pt}
	{\includegraphics[width=0.16\linewidth]{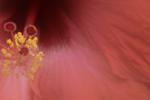}}\hspace*{-1pt}
	{\includegraphics[width=0.16\linewidth]{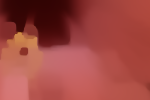}}\hspace*{-1pt}
	{\includegraphics[width=0.16\linewidth]{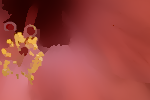}}\hspace*{-1pt}
	{\includegraphics[width=0.16\linewidth]{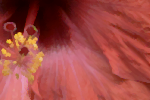}}
	
	\caption{Edge preserving smoothing results. From left to right: original, Domain Transform~\cite{DomainTransform}, Guided Filter~\cite{he2010guided}, Relative Total Variation~\cite{Xu2012}, static/dynamic filter~\cite{Ham2018} and our filter. For our filter, the iteration number is set to 10.}
	\label{fig:compare}
\end{figure*}

\begin{figure*}[!htb]
	\centering
	\begin{overpic}
		[width=0.16\linewidth]{images/compare/li.png}  
		\thicklines
		\put(55,45){\color{red}\vector(-1,-1){20}}
	\end{overpic}
	\begin{overpic}
		[width=0.16\linewidth]{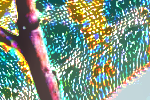}  
		\thicklines
	    \put(55,45){\color{red}\vector(-1,-1){20}}
	\end{overpic}
\begin{overpic}
	[width=0.16\linewidth]{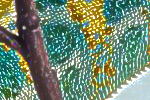}  
	\thicklines
	\put(55,45){\color{red}\vector(-1,-1){20}}
\end{overpic}
\begin{overpic}
	[width=0.16\linewidth]{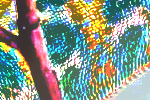}  
	\thicklines
	\put(55,45){\color{red}\vector(-1,-1){20}}
\end{overpic}
\begin{overpic}
	[width=0.16\linewidth]{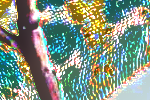}  
	\thicklines
	\put(55,45){\color{red}\vector(-1,-1){20}}
\end{overpic}
\begin{overpic}
	[width=0.16\linewidth]{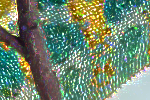}  
	\thicklines
	\put(55,45){\color{red}\vector(-1,-1){20}}
\end{overpic}

\vspace{1mm}
\begin{overpic}
	[width=0.16\linewidth]{images/compare/flower.png}  
	\thicklines
	\put(70,30){\color{green}\vector(-1,0){20}}
\end{overpic}
	\begin{overpic}
		[width=0.16\linewidth]{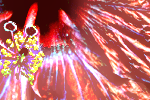}  
		\thicklines
		\put(70,30){\color{green}\vector(-1,0){20}}
	\end{overpic}
\begin{overpic}
	[width=0.16\linewidth]{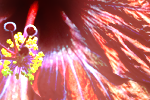}  
	\thicklines
	\put(70,30){\color{green}\vector(-1,0){20}}
\end{overpic}
\begin{overpic}
	[width=0.16\linewidth]{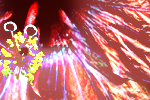}  
	\thicklines
	\put(70,30){\color{green}\vector(-1,0){20}}
\end{overpic}
\begin{overpic}
	[width=0.16\linewidth]{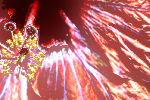}  
	\thicklines
	\put(70,30){\color{green}\vector(-1,0){20}}
\end{overpic}
\begin{overpic}
	[width=0.16\linewidth]{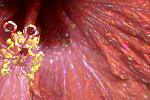}  
	\thicklines
	\put(70,30){\color{green}\vector(-1,0){20}}
\end{overpic}
	\caption{Enhancement results. From left to right: original, Domain Transform~\cite{DomainTransform}, Guided Filter~\cite{he2010guided}, Relative Total Variation~\cite{Xu2012}, static/dynamic filter~\cite{Ham2018} and our filter. The details are amplified by ten. The red or green arrows indicate artifacts.}
	\label{fig:enhance}
\end{figure*}

\begin{figure*}[!htb]
	\centering
	{\includegraphics[width=0.16\linewidth]{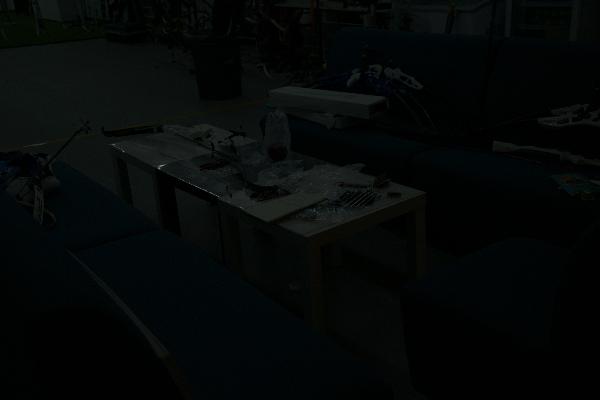}}\hspace*{-1pt}
	{\includegraphics[width=0.16\linewidth]{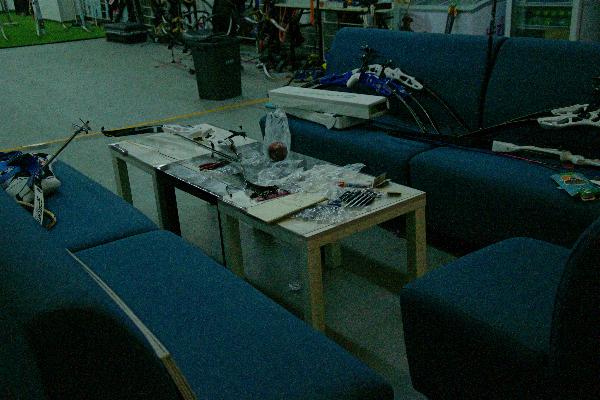}}\hspace*{-1pt}
	{\includegraphics[width=0.16\linewidth]{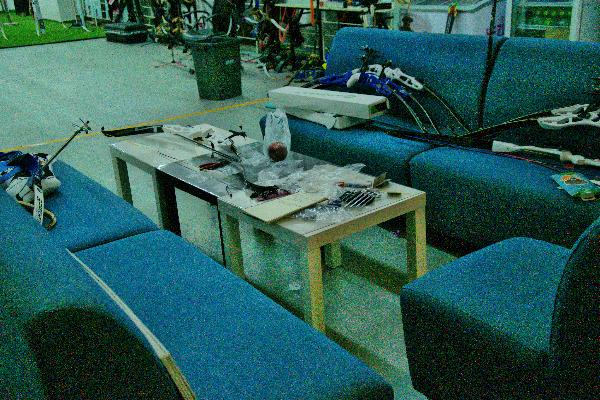}}\hspace*{-1pt}
	{\includegraphics[width=0.16\linewidth]{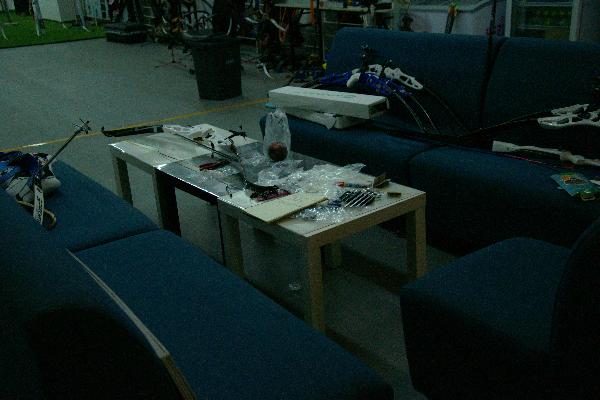}}\hspace*{-1pt}
	{\includegraphics[width=0.16\linewidth]{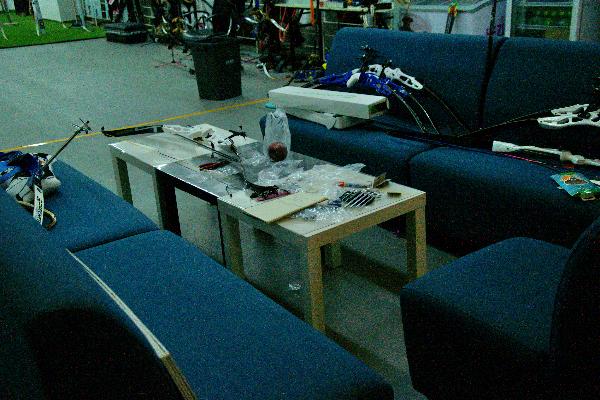}}\hspace*{-1pt}
	{\includegraphics[width=0.16\linewidth]{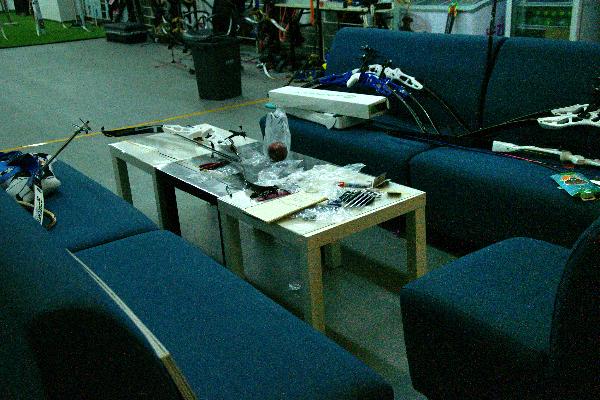}}\\
	\vspace{2pt}
	
	{\includegraphics[width=0.16\linewidth]{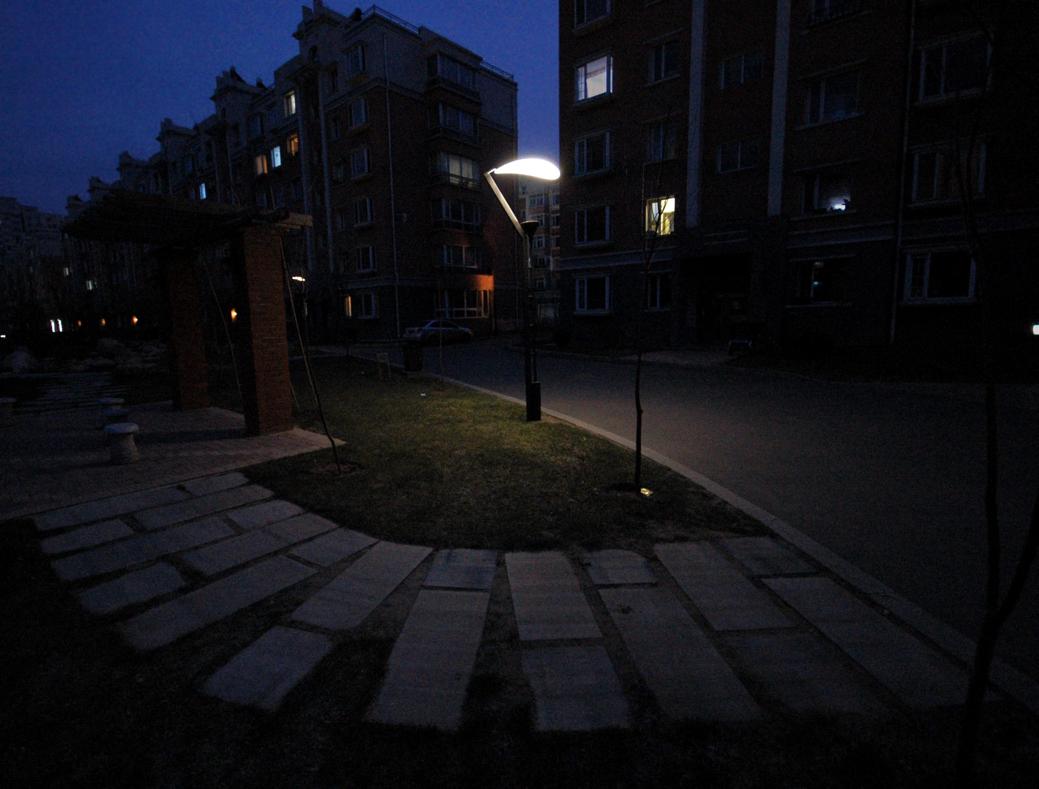}}\hspace*{-1pt}
	{\includegraphics[width=0.16\linewidth]{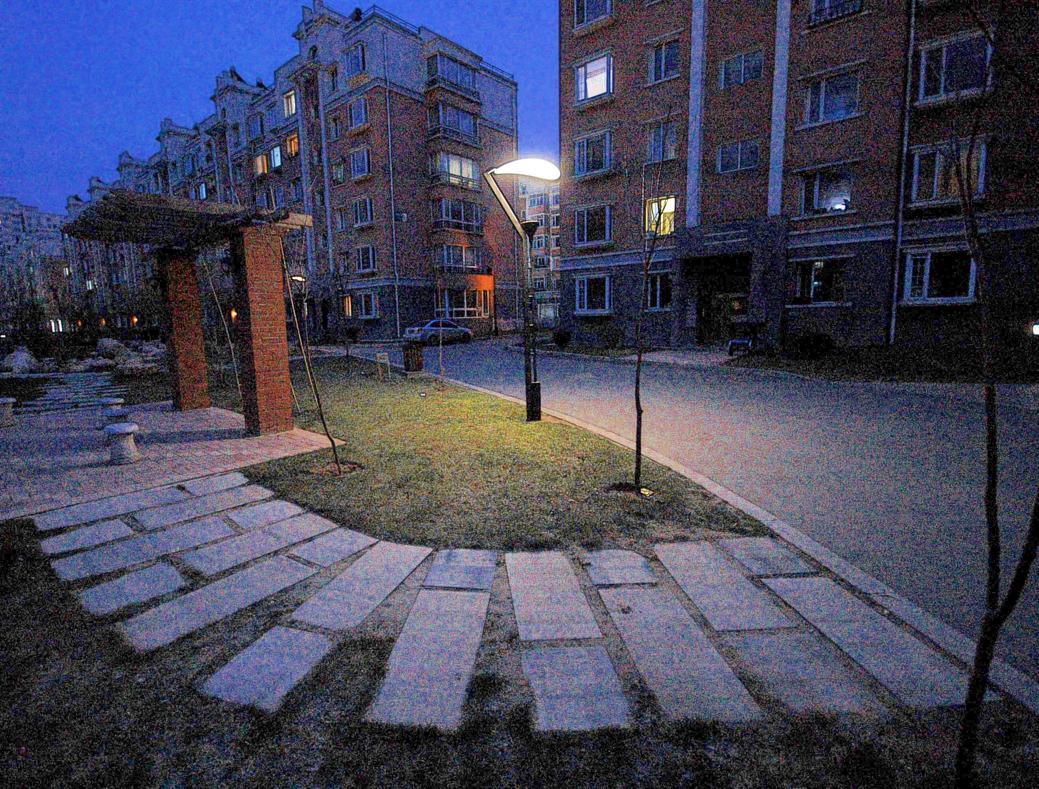}}\hspace*{-1pt}
	{\includegraphics[width=0.16\linewidth]{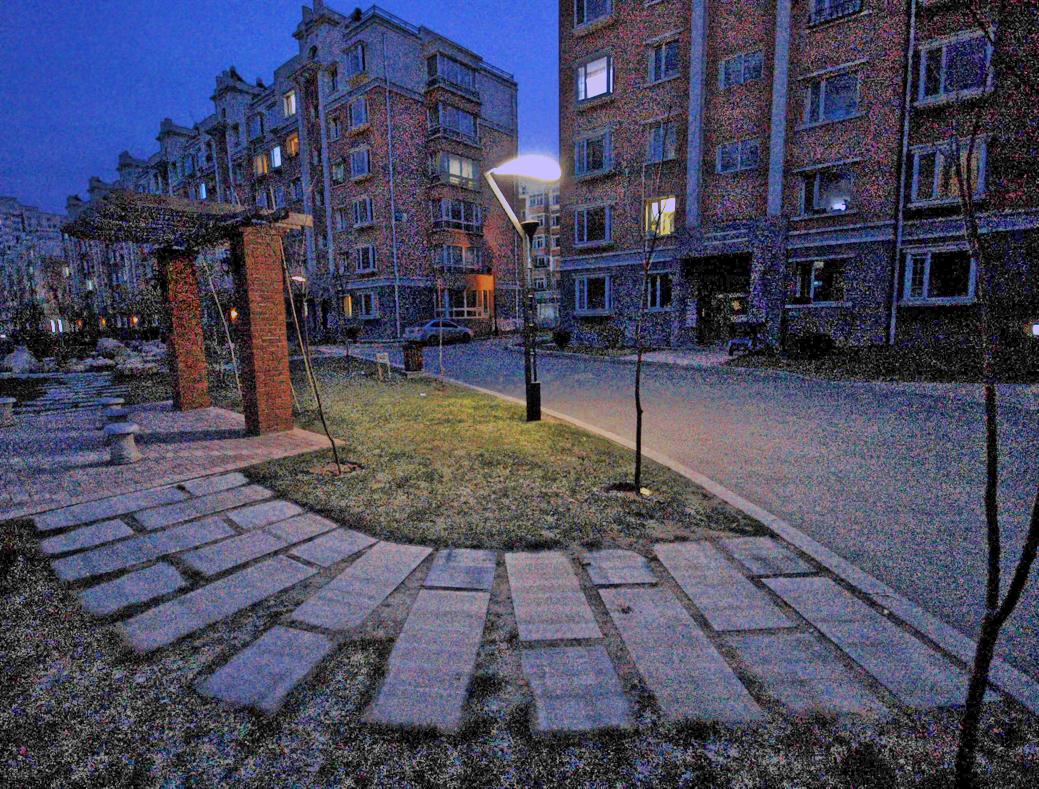}}\hspace*{-1pt}
	{\includegraphics[width=0.16\linewidth]{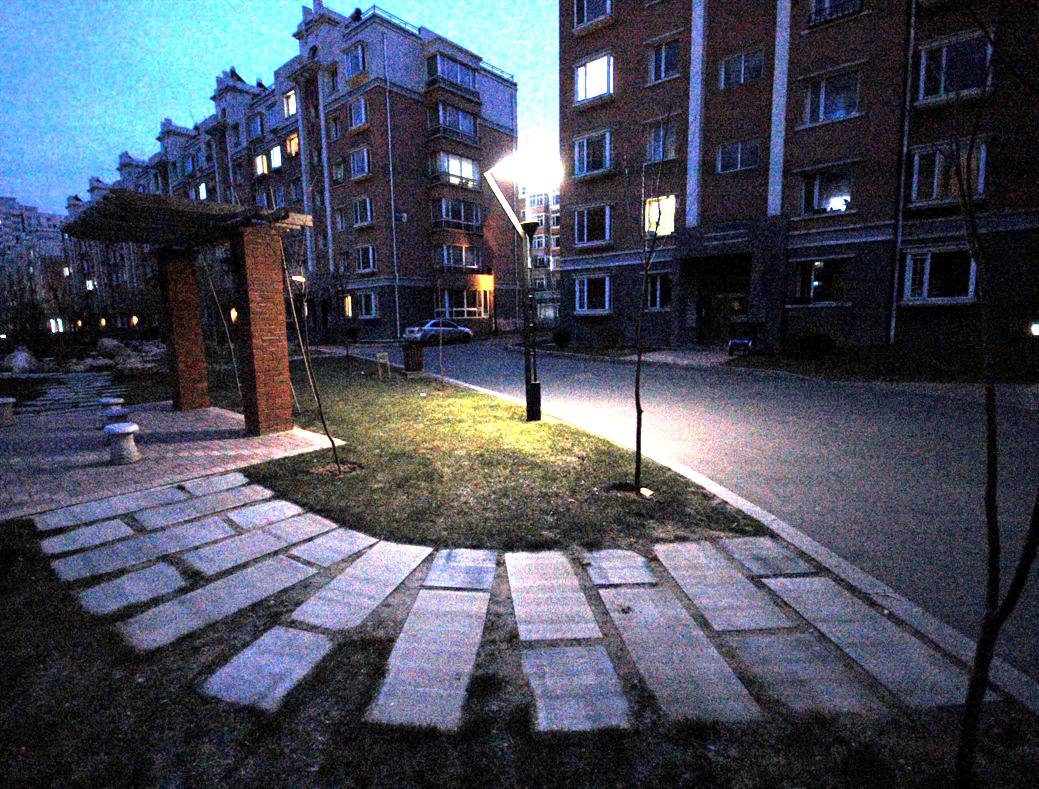}}\hspace*{-1pt}
	{\includegraphics[width=0.16\linewidth]{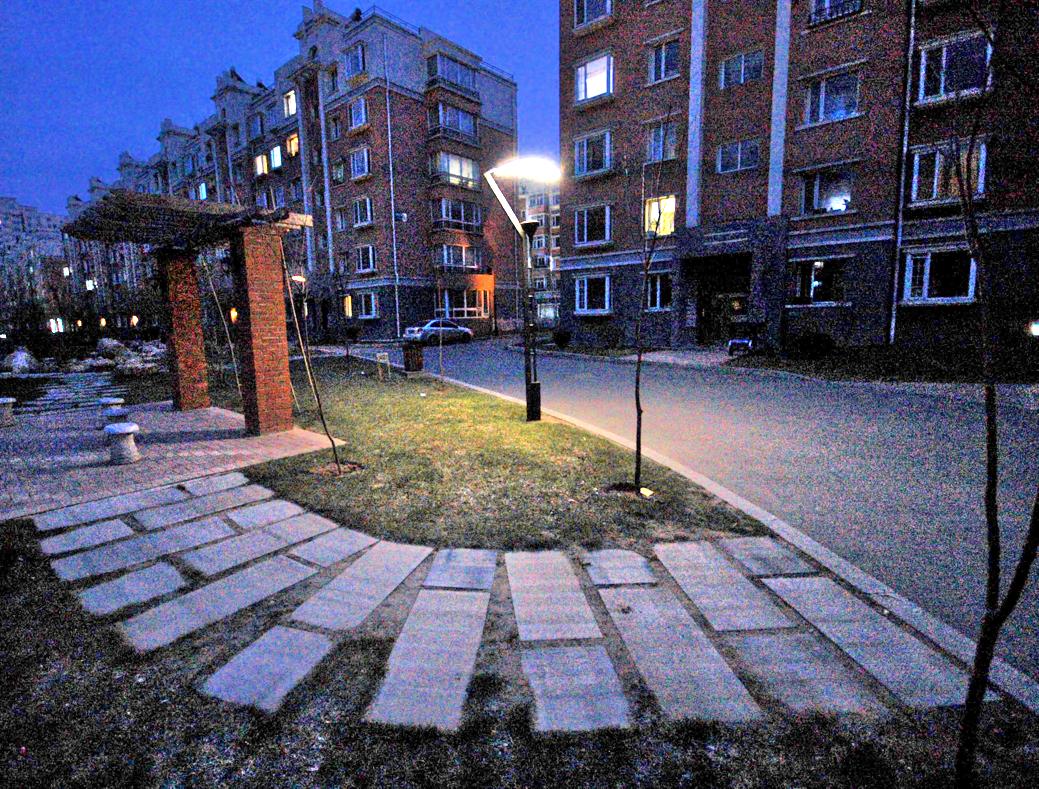}}\hspace*{-1pt}
	{\includegraphics[width=0.16\linewidth]{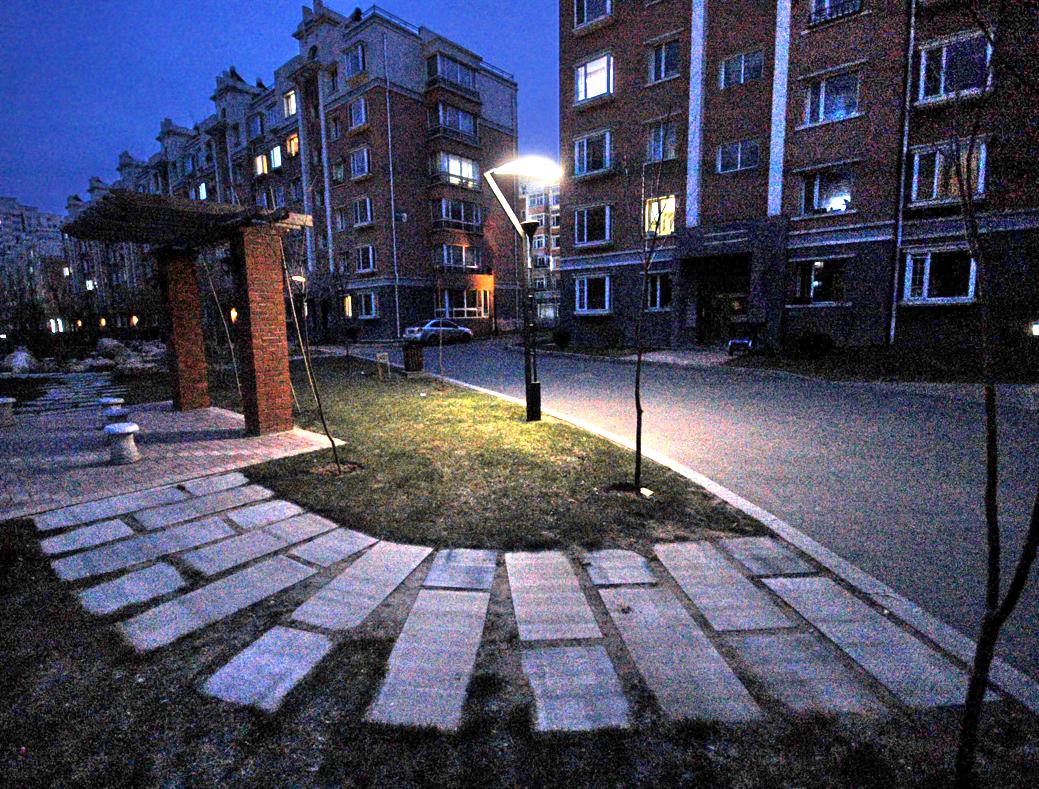}}

	\caption{Low-light image enhancement. Left to right: original, dehazing-based~\cite{dehaze}, NPE~\cite{NPE}, SRIE~\cite{SRIE}, ALSM~\cite{ALSM} and ours.}
	\label{fig:hdr}
\end{figure*}

\section{Applications}
In this section, we show several applications of the proposed filter, including image smoothing, enhancement, low light image enhancement, etc.

First, the proposed quarter Laplacian filter is compared with several edge preserving methods on color images. Our method is performed on each color channel separately. The iteration number in our method is set to 10. Other methods use their default parameters. The results are shown in Fig.~\ref{fig:compare}.

Second, we can amplify the texture details to enhance the image. The results are shown in Fig.~\ref{fig:enhance}. Some artifacts are indicated by red or green arrows. Such artifacts also indicate that these methods do not properly preserve the edges during image smoothing task. Our method has less artifacts.

Third, we use the quarter Laplacian filter to obtain details at different scales for the illumination and adjust them to enhance the light condition. The results are shown in Fig.~\ref{fig:hdr}.

\section{Conclusion}
This paper presents a quarter Laplacian filter that can smooth image with edge preserving. Its support region is $2\times2$ and it is more local. This filter can be implemented via the classical box filter, leading to high performance. It is applied in several image processing tasks, including image smoothing, texture enhancement and low-light image enhancement. It can be used in a wide range of applications~\cite{Gong:2014a,YIN2019315,Gong2019}.
\bibliographystyle{IEEEbib}
\bibliography{IP}

\end{document}